\documentclass[aps,twocolumn,showpacs,amsmath]{revtex4-1}

\usepackage[pdftex]{graphicx}
\usepackage{newlfont}
\usepackage{amssymb}
\usepackage{amsfonts}
\usepackage{amsmath}
\usepackage{keyval}
\usepackage{graphicx}
\usepackage{bm}

\usepackage{float}
\usepackage{epsfig}
\usepackage{epstopdf}
\usepackage{upgreek}

\begin{document}

\title{Monogamy of quantum correlations reveals frustration in a quantum Ising spin system: Experimental demonstration}

\author{K. Rama Koteswara Rao$^1$, Hemant Katiyar$^2$, T. S. Mahesh$^2$, Aditi Sen(De)$^3$, Ujjwal Sen$^3$, and Anil Kumar$^1$}

\affiliation{$^1$Centre for Quantum Information and Quantum Computation,\\
 Department of Physics and NMR Research Centre, Indian Institute of Science, Bangalore 560012, India\\
 $^2$Department of Physics and NMR Research Center, Indian Institute of Science Education and Research, Pune 411008, India\\
 $^3$Harish-Chandra Research Institute, Chhatnag Road, Jhunsi, Allahabad 211 019, India}

\begin{abstract}
We report a nuclear magnetic resonance experiment, which simulates the 
quantum transverse Ising spin system in a triangular configuration and further show that the monogamy of 
quantum correlations can be used to distinguish between the frustrated and non-frustrated regimes in the ground state of this system. Adiabatic state preparation methods are used to prepare the ground states of the spin system. 
We employ two different multipartite quantum correlation measures to analyze the experimental ground state of the system in both the frustrated and non-frustrated regimes. 
In particular, we use multipartite quantum correlation measures generated by monogamy considerations of negativity, a bipartite entanglement measure, and that of quantum 
discord, an information-theoretic quantum correlation measure. As expected from theoretical predictions, 
the experimental data confirm that the non-frustrated regime shows higher multipartite quantum correlations compared
 to the frustrated one.
\end{abstract}


\maketitle

\section{Introduction}

Recently, quantum correlations\cite{HHHHRMP, discord-rmp}
have  been employed beyond its traditional ambit of quantum computation and information, and 
used as tools in  
quantum many-body physics\cite{ADUSadv, FazioRMP, discord-rmp}.
Detecting phases of many-body systems is an important task with fundamental as well as technological implications \cite{phase-tran}. 
Frustrated cooperative phenomena form one of the centerstages of the research in many-body physics. Frustration in a many-body Hamiltonian appears 
when it is not possible to simultaneously minimize each of the energy bonds in the Hamiltonian independently, and may appear as a result of competing interactions 
due to the geometry of the lattice 
or due to incommensurate values of the coupling strengths \cite{frustu-gen-ref}. Frustrated systems usually possess hugely degenerate ground states and a rich phase 
diagram ranging from quantum spin liquids to resonating valence-bond states  \cite{frustu-rich-phase-diagram}. Very recently, theoretical studies have shown that 
entanglement can be an effective tool for investigating frustrated quantum systems \cite{amaderPRL, IlluminatoPRL}. Frustrated interactions had been known to be 
present in several solid state systems \cite{frustu-rich-phase-diagram}. Recent experimental breakthroughs have made it possible to engineer frustrated 
spin models in ultracold atoms in optical lattices \cite{expoptical}, trapped ions \cite{expions},  NMR \cite{expNMR}, etc, and have led to the 
 possibility of 
observing the effects of entanglement in the different phases of frustrated spin models in the laboratory.

In the present work, we employ multipartite quantum
correlations to  distinguish the frustrated regimes from the non-frustrated ones.
There are many ways by which multipartite quantum correlations  can be quantified \cite{HHHHRMP,discord-rmp} 
and even though their properties can vary widely, there are some distinct connecting 
themes. One of them is the ``monogamy'' of bipartite quantum correlations, 
which 
broadly demands that if two parties are strongly quantum correlated, they cannot have a significant amount of the same with a third party  
\cite{Ekert, Bennett, CKW}.

There is a two-fold aim of the present work. First of all, we want to experimentally observe the effect of monogamy of bipartite 
quantum correlations in a multiparty quantum system.  Secondly, we wish to apply the concept of monogamy of such correlations to distinguish between 
frustrated and non-frustrated regimes in a quantum Ising spin system. 
To attain our goal, we prepare a transverse Ising spin system \cite{TIM-book} in a triangular 
configuration by using NMR techniques. 
It undergoes a transition from a non-frustrated regime to a frustrated one when its coupling strength is varied from a ferromagnetic to an antiferromagnetic regime.
We study the multipartite quantum correlations  of the ground state to detect the two different regimes.   
Specifically, we consider multiparty quantum correlation measures, generated from monogamy studies of bipartite quantum correlations
 \cite{Ekert, Bennett, CKW}.  
In analyzing the experimentally generated ground state, 
we employ 
monogamy  of (i) negativity \cite{CKW, logneg, monogamy-other},  which is a bipartite entanglement measure, and of (ii) quantum discord \cite{discord,discordmonoama}, which is an 
information-theoretic  quantum correlation measure. 
For such investigations, we initially prepare the spin system in the ground state of the transverse field Hamiltonian, which is a product state and then
 adiabatically drive it to both frustrated and non-frustrated regimes in such a way that the system remains in the ground state of the instantaneous 
Hamiltonian. We then calculate the multipartite correlations of the ground state by performing quantum state tomography at different time intervals.
Moreover,  coinciding with the theoretical simulations, we observe that the non-frustrated regime has higher multipartite quantum correlations than the frustrated one. 
The transition point from the non-frustrated 
to the frustrated regime is well indicated by the vanishing monogamy relation.

The paper is arranged as follows. In Sec. \ref{sec-measure}, we present the  multipartite quantum correlations that are employed in analyzing the ground state. 
We describe the frustrated and non-frustrated quantum Ising spin systems in Sec. \ref{fIsing}.
The  experimental results are presented in Sec. \ref{sec-results} and we conclude in Sec. \ref{sec-conclu}.

\section{Multipartite Quantum Correlations}
\label{sec-measure}

Quantum correlations of a multipartite quantum state can be quantified in a variety of approaches. 
A prominent one among them is to use the concept of monogamy of bipartite quantum correlations \cite{CKW, monogamy-other, discordmonoama}. 
In a tripartite scenario, monogamy of a bipartite quantum correlation restricts the amount of that correlation which
 can be shared between the three parties. Let us suppose 
that  three parties, Alice, Bob, and Charu, denoted respectively as \(1\), \(2\), and  \(3\), share a tripartite quantum state. 
Monogamy of quantum correlations implies that if \(1\)  has substantial quantum correlations with \(2\), it can have only a restricted amount of the same with 
\(3\).  To state it more precisely, for a bipartite quantum correlation measure \(\mathcal{Q}\) and a tripartite quantum state \(\rho_{123}\), 
one can introduce a quantity, known as ``monogamy score for \(\mathcal{Q}\)'' \cite{ManabPRA} and denote as \(\delta_{\mathcal{Q}}\), given by  
\begin{equation}
\delta_{\cal Q}(\rho_{123}) = {\cal Q}_{1(23)} - {\cal Q}_{12} - {\cal Q}_{13},
\label{monogen}
\end{equation} 
where \(\mathcal{Q}_{1(23)}\) is the bipartite quantum correlation measured for the state \(\rho_{123}\) in the \(1:23\) partition. 
\(\mathcal{Q}_{12}\) is the same measure for the reduced state \(\rho_{12}=\mbox{tr}_3\rho_{123}\), and 
similarly for \(\mathcal{Q}_{13}\). 
The three-party quantum state \(\rho_{123}\) is said to be  monogamous with respect to the bipartite quantum correlation measure \(\mathcal{Q}\), 
if \(\delta_{\cal Q}(\rho_{123}) \geq 0\). 
A measure for which
all  tripartite quantum states produce  non-negative monogamy scores  is  said to be ``monogamous''.

The monogamy score, and its properties, will certainly depend on the bipartite quantum correlation measure that is employed to define the score. There are many ways in which one can conceptualize bipartite quantum correlations, and 
correspondingly there are many bipartite quantum correlation measures \cite{HHHHRMP, discord-rmp}. Broadly, these measures are defined within two paradigms, viz. the entanglement-separability paradigm and the information-theoretic one. 
We consider bipartite quantum correlation measures from both these paradigms to define monogamy scores, that are subsequently used for distinguishing frustrated regimes of quantum spin systems from non-frustrated ones in the experiment.
As we will see below, one of these measures is  monogamous for pure three-qubit quantum states, while the other is not.

\emph{Monogamy Score for Negativity Squared.} -- Negativity is an important quantum correlation measure for two-party quantum states \cite{logneg}. 
It is defined within the entanglement-separability paradigm.
The negativity, \(N_{12}\), of an arbitrary bipartite quantum state $\rho_{12}$ is the
absolute value of the sum of the negative eigenvalues of the
partial transposed state \(\rho_{12}^{T_{1}}\), where the partial transposition is taken with respect to Alice \((1)\).
The importance of this measure arises from the fact that if a partially transposed bipartite quantum state has negative eigenvalues, the state must be entangled 
\cite{Peres_Horodecki}.

Replacing \({\cal Q}\) in Eq. (\ref{monogen}) by the squared  negativity, we obtain the monogamy score for the negativity squared,   given by
\begin{eqnarray}
 \delta_{N^2}=N^2_{1(23)}-N^2_{12}-N^2_{13}.
 \label{mononeg}
\end{eqnarray} 
Recently, it has been shown  that the negativity squared  is monogamous, 
i.e. \(\delta_{N^2} \geq 0\), for all 
three-qubit pure states \cite{monogamy-other}. 
In this paper, we measure \(\delta_{N^2}\) in both the frustrated and non-frustrated regimes. For ease of reference, we call the monogamy score for negativity squared as the ``entanglement monogamy score''.

\emph{Quantum Discord and Monogamy.} -- Let us now define another bipartite  measure of quantum correlation, and importantly it
 does not belong to the entanglement-separability paradigm. 
Quantum discord is defined as the difference between two classically equivalent formulations of mutual information, when the systems involved are quantum \cite{discord}, and 
is given by
\begin{equation}
D_{12} = D(\rho_{12})=I(\rho_{12})-J(\rho_{12}),
\label{disc}
\end{equation}
where \(I(\rho_{12})\) and \(J(\rho_{12})\) are argued to be, respectively,  measures of total and classical correlations of  \(\rho_{12}\). 
\(I(\rho_{12})\) is defined as \(S(\rho_1)+S(\rho_2)-S(\rho_{12})\), where \(S(\sigma) = -$tr$ (\sigma \log_2 \sigma)\) 
is the von Neumann entropy of \(\sigma\), \(\rho_1 = $tr$_2 \rho_{12}\),  and  \(\rho_2 = $tr$_1 \rho_{12}\).
\(J(\rho_{12})= S(\rho_{2})-S(\rho_{2|1})\), where  
the conditional entropy \(S(\rho_{2|1})= \min_{{\Pi_i^1}} \sum_i p_i S(\rho_{2|\Pi_i^1})\), with the minimization being performed over all possible 
rank-one projection-valued measurements \({\Pi_i^1}\) 
on subsystem 1. Here the output state \(\rho_{2|\Pi_i^1}=\Pi_i^1 \rho_{12} \Pi_i^1 / $tr$_{12}(\Pi_i^1 \rho_{12})\), and the probability \(p_i=$Tr$(\Pi_i^1 \rho_{12})\). The monogamy score for quantum discord (referred later as the ``discord monogamy score'')
 for a tripartite state \(\rho_{123}\) is given by 
\cite{discordmonoama} 
\begin{equation}
\delta_{D}=D_{1(23)}-D_{12}-D_{13}.
\label{monodisc}
\end{equation}
Unlike the entanglement monogamy score, the discord monogamy score, \(\delta_{D}\), can be both nonnegative and negative, even for pure tripartite states \cite{discordmonoama}. 

\section{Frustrated Ising spin system}
\label{fIsing}

Frustrated spin sytems have attracted a lot of interest due to their rich phase diagrams. 
Frustration  can  be observed in a system consisting of three quantum spin-1/2 particles positioned at the corners of an equilateral traingle, 
and having Ising (nearest-neighbor) interactions. The Hamiltonian of this three-spin transverse 
Ising model is given by
\begin{equation}
\cal{H}=h(\sigma^1_x+\sigma^2_x+\sigma^3_x)+J(\sigma^1_z \sigma^2_z+\sigma^2_z \sigma^3_z+\sigma^1_z \sigma^3_z), 
\end{equation}
where \(h\) is the strength of the transverse field, \(J\) is the coupling strength of the  Ising interactions, and \(|J| \gg h\). The three quantum 
spin-1/2 particles are denoted as \(1,2,3\). \(\sigma_x^i\) and  \(\sigma_z^i\),
 for \(i=1,2,3\), are the Pauli spin matrices at site \(i\).
When \(J>0\), i.e, the Ising interactions are of anti-ferromagnetic type, the system is frustrated, whereas when \(J<0\), i.e, the Ising interactions are of 
ferromagnetic type, the system is non-frustrated.

We now describe the adiabatic state preparation method used to prepare the ground state of this spin system in the laboratory \cite{adiabatic}. The quantum adiabatic theorem  
states that if a system is initially in the ground state and if its Hamiltonian evolves slowly with time, it will be found at any later time in the ground state of 
the instantaneous Hamiltonian \cite{Messiah}. 
The Hamiltonian evolution rate is governed by the relation
\begin{equation}
\label{adiacond}
\frac{| \langle 1;t| \frac{d{\cal H}(t)}{dt} |0;t \rangle |}{g^2(t)} \le \epsilon,
\end{equation}
\noindent where \(\epsilon\) is a small number, \(|0;t\rangle\) and \(|1;t \rangle\) are respectively the ground and first excited states of the 
instantaneous Hamiltonian \({\cal H}(t)\), and \(g(t)\) is the energy difference between the corresponding energy levels. The system stays in the instantaneous 
ground state of the Hamiltonian with a probability \( (1-\epsilon^2)^2 \).

\begin{figure}
 \centering
 \includegraphics[scale=0.6]{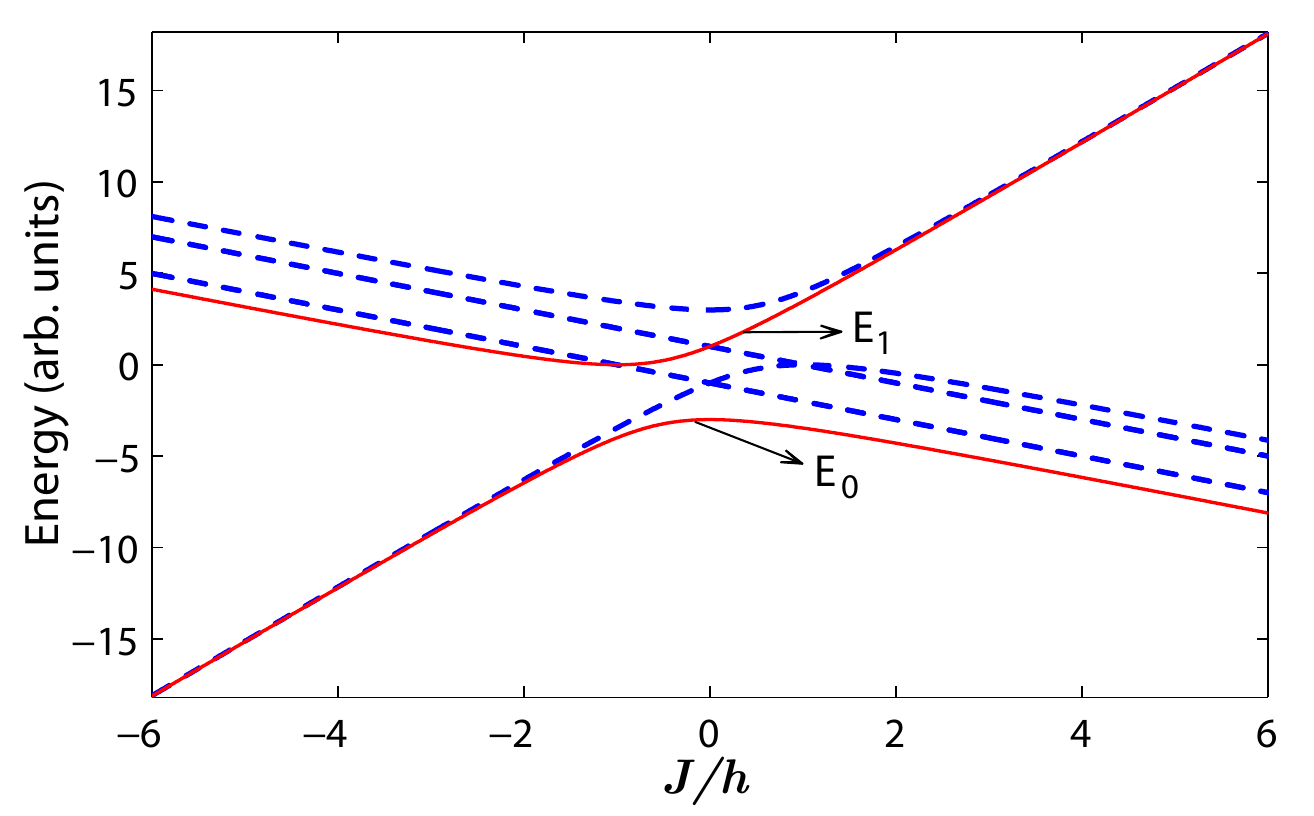}
 \caption{Energy level diagram. The red solid curves, which are marked as \(E_0\) and \(E_1\), represent respectively the energy levels of the ground state and the excited state which is relevant in the calculation of the adiabatic evolution 
rate (Eq. (\ref{adiacond})). The blue dashed curves represent the energy levels corresponding to the other excited states.}
 \label{fig1} 
\end{figure}

The spin system is initially prepared in the ground state of the Hamiltonian \(h(\sigma_x^1+\sigma_x^2+\sigma_x^3)\). Then the system is taken to the frustrated 
regime by adiabatically increasing \(J\) from \(0\) to \(|J_{max}|\) and similarly to 
the non-frustrated regime by decreasing \(J\) from \(0\) to \(-|J_{max}|\), where \(|J_{max}|\gg h\). The system stays in the ground state of the instantaneous 
Hamiltonian with high probability, if \(J\) is changed slowly enough  
so that it satisfies Eq. (\ref{adiacond}). The energy level diagram  of the spin system is shown in the Fig. \ref{fig1}. The energy level of the ground state of the 
system is represented by the solid red curve, which is marked as \(E_0\). 
The energy level of the only excited state which is relevant in the calculation of the adiabatic evolution rate of the Hamiltonian is represented by the other solid red 
curve, which is marked as \(E_1\).
The energy levels of all the other excited states are shown by blue dashed curves. Though there are energy levels in between \(E_0\) and \(E_1\), there are no possible transitions 
from the ground state to these excited states, as the 
transition amplitudes (given by the matrix elements similar to the numerator of the left hand side in  Eq. (\ref{adiacond})) are zero in these cases.

\section{Experimental Implementation}
\label{sec-results}

\begin{figure}
 \includegraphics[scale=0.45]{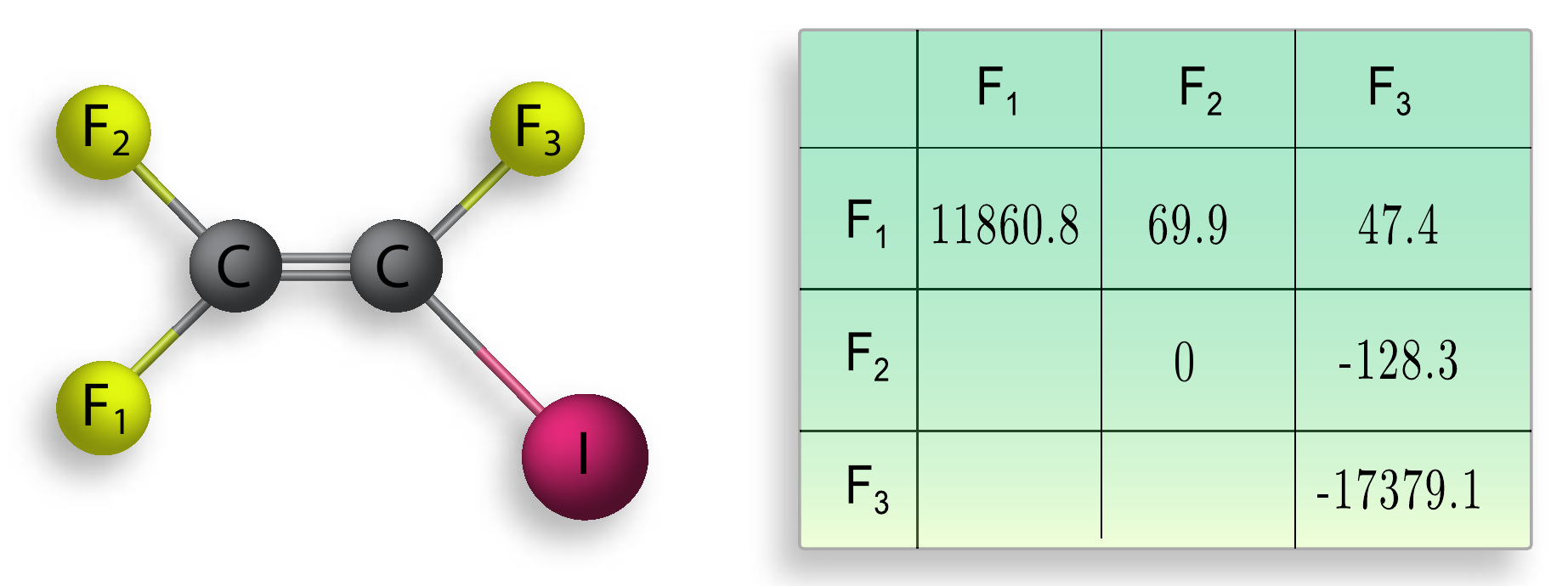}
 \caption{Chemical structure of the molecule (left) and the table of Hamiltonian parameters (right). 
 In the table, the diagonal elements are the chemical shifts \(\nu_i\) (in Hz) of the fluorine spins and the off-diagonal elements are the 
scalar coupling constants \(J_{ij}\) (in Hz) between them.}
 \label{fig2}
\end{figure}

The spin-system chosen for the experiments is 
iodotrifluoroethylene \((\mathrm{C_2F_3I})\) dissolved in acetone-D$_6$.
Here the three $^{19}$F nuclear spins act as three spin-1/2 particles.
The chemical structure of the molecule, the chemical shifts of the three fluorine nuclei, and the \(J\)-couplings between them are shown in Fig. \ref{fig2}. 
The experiments have been carried out at a temperature of 290 K in an 11.7 Tesla magnetic field on a Bruker UltraShield AV III 500 MHz NMR spectrometer using a QXI probe. 
The \(\mathrm{^{19}F}\) resonance frequency at this field is 470 MHz.
The measured longitudinal relaxation time constants ($\mathrm{T_1}$) of the three fluorine nuclei $\mathrm{F_1}$, $\mathrm{F_2}$, and $\mathrm{F_3}$ are 6.9 s, 7.5 s, and 6.2 s respectively. The transverse relaxation time constants ($\mathrm{T_2}$) of these nuclei, measured by Hahn echo, are 2.8 s, 3.1 s, and 3.3 s respectively.

In the rotating frame, the NMR Hamiltonian of a weakly coupled three-spin system is given by
\begin{eqnarray}
{\cal H}_\mathrm{NMR}=-\sum_{i=1}^3\pi\nu_i\sigma^i_z + \sum_{i<j,=1}^3 \frac{\pi}{2} J_{ij}\sigma^i_z\sigma^j_z.
\end{eqnarray}
\noindent where \(\nu_i\) are the chemical shifts of the \(\mathrm{^{19}F}\) nuclear spins, \(J_{ij}\) are the scalar coupling constants between them.
The equilibrium density matrix under high temperature and high field approximation, is in a highly mixed state, given by
\begin{equation}
\rho_{eq}=\tfrac{1} {8}(I+\zeta \ \Delta \rho_{eq}),
\end{equation}
where $\zeta \sim 10^{-5}$ is the purity factor and  the deviation part of
the density matrix \cite{Ernst}
\begin{equation}
\Delta\rho_{eq}\propto\sigma^1_z+\sigma^2_z+\sigma^3_z.
\end{equation}
In liquid state room temperature NMR, since the preparation of a pure state 
requires extreme conditions,
it is a common practice to prepare a pseudo-pure state (PPS) that mimics the pure state \cite{PPS1,PPS2}. 
We have used the spatial averaging method to prepare the \(|000\rangle\) PPS from the equilibrium \cite{PPS1}.
The total time taken to prepare this PPS is $\approx$ 30 ms.

In the experimental implementation, the Hamiltonian \(\cal H(t)\) is discretized into M+1 steps as \(\cal H\left(\frac{m}{M}T\right)\), where \(T\) 
is the total duration of the adiabatic evolution and \(m\) goes from 0 to \(M\). 
The unitary operator corresponding to the each step is given by
\begin{eqnarray}
U_m= \exp(-i\cal H(\tfrac{m}{M}T)\Delta t)= \exp(-i[h(\sigma^1_x+\sigma^2_x+\sigma^3_x)\nonumber\\
+J(\tfrac{m}{M}T)(\sigma^1_z \sigma^2_z+\sigma^2_z \sigma^3_z+\sigma^1_z \sigma^3_z)]\Delta t),\nonumber\\
\end{eqnarray}
\noindent where \(\Delta t=\frac{T}{M+1}\). The unitary operator corresponding to the total adiabatic evolution after the \((M+1)\)th step can be written as
\begin{equation}
\label{combunit}
U=\prod^M_{m=0} U_m.
\end{equation}
The unitary operator for each step can be approximated to second order in $\Delta t$ by using the Trotter's formula
\begin{eqnarray}
\label{trotaprx}
U_m \approx \ &&\exp(-ih(\sigma^1_x+\sigma^2_x+\sigma^3_x)\tfrac{\Delta t}{2}) \nonumber\\ &&\times \exp(-iJ(\tfrac{m}{M}T)(\sigma^1_z \sigma^2_z+\sigma^2_z \sigma^3_z+\sigma^1_z \sigma^3_z)\Delta t) \nonumber\\ && \times \exp(-ih(\sigma^1_x+\sigma^2_x+\sigma^3_x)\tfrac{\Delta t}{2}).
\end{eqnarray}
In the experiment, the value of \(h \Delta t\) was set to \(\pi/21\) and that of \(J(T) \Delta t\) to \(\pi/4\) and \(-\pi/4\) in the frustrated and non-frustrated cases respectively. 
The value of \(\frac{m}{M}\) was increased from 0 to 1 in 21 steps, in both the frustrated and non-frustrated cases. 
Considering the energy gap between the ground state ($E_0$) and the relevant excited state ($E_1$), increasing \(J(t)\) linearly is not time-efficient. To achieve a time-efficient
adiabatic evolution, we used a sine hyperbolic variation in \(J(t)\). 
By using experimental parameters, we have estimated $\epsilon$ from the adiabatic relation of Eq. (\ref{adiacond}) and its maximum value is 0.063. This corresponds to a minimum probability of 0.992 for the system to stay in the ground state.

The ground state of the Hamiltonian at time \(t=0\), 
i.e. of $h(\sigma^1_x+\sigma^2_x+\sigma^3_x)$, is $ |---\rangle $ with \(|-\rangle=\tfrac{1}{\sqrt{2}}(|0\rangle-|1\rangle)\). This state was prepared from the 
\(|000\rangle\) PPS by applying a \(\frac{\pi}{2}\) rotation with respect to the \(-y\) axis on
all the three spins. This rotation was realized by a numerically optimized amplitude and phase modulated radio frequency (RF) 
pulse using GRadient Ascent Pulse Engineering (GRAPE) technique \cite{GRAPE,Ryan}. 
The length of this pulse is 600 $\mu$s.

\begin{figure*}

 \includegraphics[width=16cm]{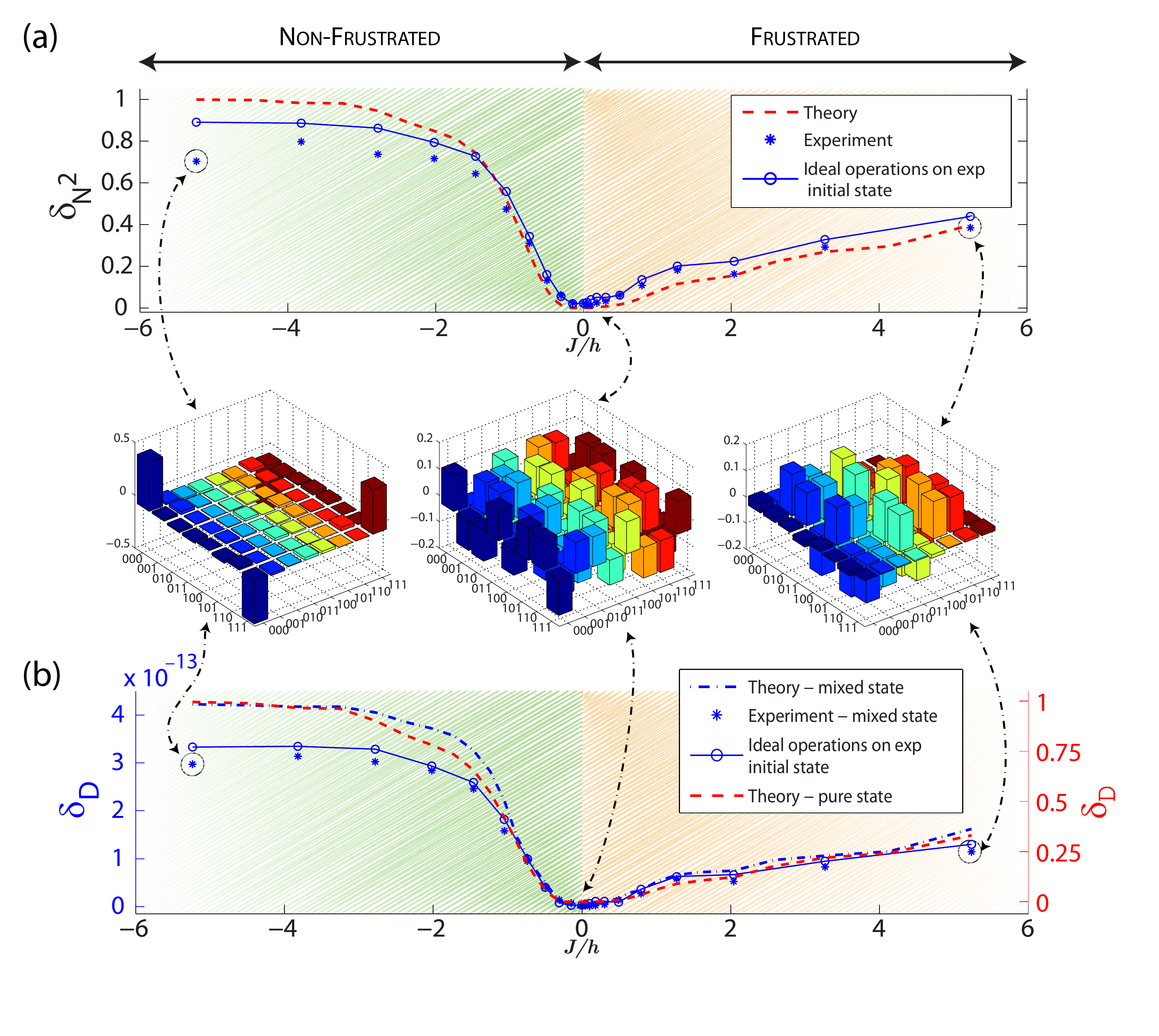}
 \caption{(a) Entanglement and (b) Discord monogamy scores. 
 The red-dashed curves correspond to the theoretically expected results, obtained by applying ideal unitary operations 
 on the ideal initial state. The blue circles are obtained by applying the ideal unitary operations on the experimental initial state. 
 The blue  stars correspond to experimental results.
 In (b) the blue-dash-dotted curves correspond to the theoretically expected results for the mixed states of 
 the NMR system, and the y-axis for the red-dashed curve is given at the right.
 For simulating all
  the theoretically expected density matrices, we used the Trotter's approximation of Eq. (\ref{trotaprx}) 
 and fixed the total number of steps for the adiabatic evolution as 21 for
 each, in both the regimes.
 The real parts of the experimental density matrices are also shown 
 for the initial state \(|---\rangle \langle---|\) (middle one), the final state in 
 the non-frustrated regime (left one), and the final state in the frustrated regime (right one).}
 \label{fig3}
\end{figure*}

The $J$-couplings of the spin system are unequal and constant, which cannot be changed directly. However, with judicious use of RF pulses one can create the effective Hamiltonians to simulate the adiabatic evolution of Eqs. (\ref{combunit}) and (\ref{trotaprx}). 
Here, we created these effective Hamiltonians by modulating the amplitude and phase of the RF pulses using GRAPE algorithm.
As evident from Eq. (\ref{combunit}), the adiabatic evolution after the \(k^{\mathrm{th}}\) step can be written as a product of \(k\) unitary operators (\(U_{k}\)s) acting on the initial state. 
For efficient implementation of the adiabatic evolution, we generated GRAPE 
pulses by cascading these unitary operators.
For example, in the 5th step, the unitary operator $U=U_4 U_3 U_2 U_1 U_0$ was realized by a single GRAPE pulse and applied on the initial state  $|---\rangle$. 
This way we generated different GRAPE pulses for the combined unitary operators of the steps 3, 5, 7, $\cdots$ , 21 and applied on the initial state in both the regimes.
The length of the pulses, which were used to realize the full adiabatic evolution (for 21 steps) in both the regimes, is 30 ms. The length of the pulses for realizing all the intermediate steps ranged between 1 ms to 30 ms.
All the GRAPE pulses were optimized such that they are robust against RF field inhomogeneity and the average 
Hilbert-Schmidt fidelity of all these pulses are greater than 0.995.
Quantum state tomography of the full density matrix was performed after the steps 3, 5, 7, $\cdots$ , 21 in both the frustrated and non-frustrated regimes. The reconstruction of the full density matrix involves an optimized set of 7 experiments and 
fitting of the corresponding real and imaginary parts of the single quantum spectra of all the three \(\mathrm{^{19} F}\) nuclear spins \cite{tomo}. 
The real part of the reconstructed density matrices corresponding to the initial state \(|---\rangle \langle---|\) and that of the final states corresponding to the last step in both the regimes are shown in the Fig. \ref{fig3}.

To quantitatively evaluate the experimental results, we calculate the fidelity (\(F\)) of the experimental density matrices (\(\rho_{\mathrm{exp}}\)) with 
respect to the theoretical density matrices (\(\rho_{\mathrm{th}}\)), given by
\begin{equation}
F=\frac{\textrm{tr}(\rho_{\mathrm{th}} \rho_{\mathrm{exp}})}{\sqrt{\textrm{tr} (\rho_{\mathrm{th}}^2) \textrm{tr} (\rho_{\mathrm{exp}}^2)}}.
\end{equation}
The fidelity of the initial state \(|---\rangle \langle---|\) was found to be ~0.99 and that of all other final density matrices were greater than ~0.984.

The entanglement and discord monogamy scores were calculated by using the relations in Eq. (\ref{mononeg}) and Eq. (\ref{monodisc}) respectively. We make an important change in our procedure to calculate the quantum discord of the experimental states. 
Instead of using only the pseudo-pure part of the experimental density matrix, as was done for calculating the 
negativity,  we used the full mixed state density matrix of the NMR system  for calculating the quantum discord. 
It is well known that the room temperature liquid-state NMR systems can have non-zero discord, although  the purity of these systems is too small 
to exhibit any real entanglement \cite{Caves,Hemantdisc}. The negativity \(N_{12}\) and the quantum discord \(D_{12}\) measured from the experimental states along with the theoretically expected ones are given in the Appendix.

The experimental results along with the theoretically expected ones for entanglement monogamy score and that for discord monogamy score are shown in Fig. \ref{fig3}(a) and Fig. \ref{fig3}(b) respectively. It is clear from these results that the non-frustrated regime has higher multipartite quantum correlations compared to the frustrated one.  
In the case of discord monogamy score (Fig. \ref{fig3}(b)), the theoretically expected results are shown for both the pure (dashed red curve) and the NMR mixed state (dash-dotted blue curve) density matrices, where the latter's pseudo-pure part is proportional to the former. Although the discord monogamy scores of the mixed states are very small, their overall behaviour is very much similar to that of the pure states. To qualitatively analyse the error due to the imperfect initial state, we calculated the entanglement and discord monogamy scores by applying ideal operations on the experimental initial state and the results are also shown in the Fig. \ref{fig3}. The overall agreement of the experimental results with the theoretically expected ones that consider the experimental initial state is remarkable.
By comparing 
Figs. \ref{fig3}(a) and \ref{fig3}(b), we conclude that the monogamy scores  for negativity squared  and quantum discord have a
 similar behaviour in both the frustrated and non-frustrated regimes. This is despite the fact that 
the two corresponding bipartite quantum correlations are defined through widely different approaches -- 
one is via an entanglement-separability criterion and the other is through an information-theoretic paradigm.
As noted before, the behaviour of each one of the multiparty quantum correlation measures is different in the frustrated and non-frustrated regimes. 
Therefore, the monogamy of quantum correlations turns out to be effective in distinguishing frustrated regimes from non-frustrated ones.
\section{Conclusion}
\label{sec-conclu}
Quantum correlation of separated quantum systems are known to be useful in a variety of phenomena. However, its detection and  quantification remain difficult tasks, specially in a multipartite domain.  
In this work, we have been able to use quantum correlations to experimentally discern between frustrated and non-frustrated regimes of a 
triangular arrangement of quantum spins, in a nuclear magnetic resonance system. To attain this goal, we have used the behavior of 
multipartite quantum correlation measures of the 
ground states of this system. These multiparty measures are obtained by using the concept of monogamy of quantum correlations, which puts constraints on the sharability of quantum correlations. 
We believe the present study to be  important not only for understanding quantum correlations
in general but also in analysing various quantum phase transitions in many-body quantum systems, in particular frustrated quantum systems. 
\section*{Acknowledgements}
We thank S. S. Roy for technical help. 
The use of AV500 FTNMR spectrometers of the NMR Research Centres at IISc, Bangalore and IISER, Pune, and
funding by the Department of Science and Technology, New Delhi, are gratefully acknowledged.
This work was partly supported by the DST Projects IR/S2/PU-01/2008 and SR/S2/LOP-0017/2009.

\section*{Appendix}

\begin{figure} [h]
\includegraphics[width=8.8cm]{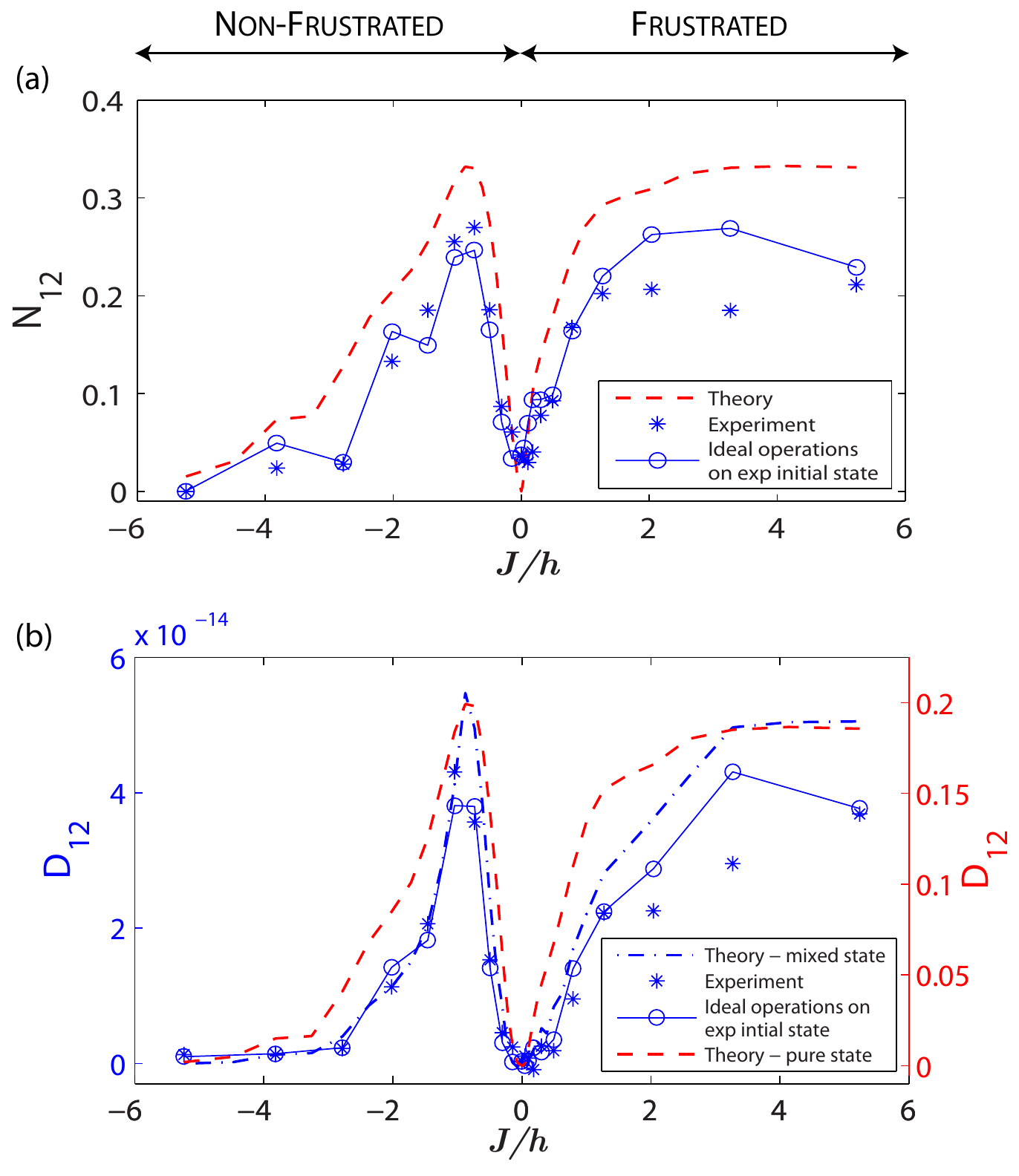}
 \caption{Bipartite quantum correlations. We plot the negativity, \(N_{12}\), in (a) and the quantum discord, \(D_{12}\), in (b).
 The red-dashed curves correspond to the theoretically expected results, obtained by applying ideal unitary operations 
 on the ideal initial state. The blue circles are obtained by applying the ideal unitary operations on the experimental initial state. 
 The blue stars correspond to the experimental results. In (b) the experimental results are calculated using the full mixed state density matrices of the NMR system and blue dash-dotted curve represent the theoretically expected results for the same, and the y-axis for the red-dashed curve is given at the right. }
 \label{fig4}
\end{figure}

The negativity \(N_{12}\) and the quantum discord \(D_{12}\) measured from the experimental states in both the frustrated and non-frustrated regimes are shown in the Fig. \ref{fig4}.
 The theoretically expected ones are also shown.
In both the regimes, both \(N_{12}\) and \(D_{12}\) increases sharply in the interval \(|J|/h\)=0 to 1. However, for higher values 
of \(|J|/h\), both the measures saturate in the frustrated regime, whereas they gradually decrease to zero in the non-frustrated 
one. 

\end{document}